\documentclass[oupthm]{my}

\usepackage{dsfont}
\usepackage{cite}
\usepackage{footnote}
\usepackage{fixfoot}
\usepackage{dsfont}
\usepackage{bigints}
\usepackage{tikz}
\usepackage{geometry}
\usepackage{enumitem}   
\newgeometry{bottom=20mm} 

\usepackage{graphicx}
\usepackage{multicol}
\usepackage{amsmath,amssymb,amsfonts, mathtools}
\usepackage{mathrsfs}

\usepackage{rotating}
\usepackage[title, toc]{appendix}
\usepackage[numbers]{natbib}
\usepackage{changepage,afterpage}
\usepackage{etruscan}
\usepackage{tikz, siunitx, newunicodechar}
\usepackage{booktabs,multirow}
\usepackage{tcolorbox}
\tcbuselibrary{skins,breakable}
\usepackage{xcolor}
\usepackage{lipsum}
\usepackage{longtable}

\usepackage{bm,bbm}
\usepackage{braket}

\usepackage[utf8]{inputenc}

\usepackage{array}
\usepackage{colortbl}

\allowdisplaybreaks[4]

\usepackage[T3,T1]{fontenc}
\DeclareSymbolFont{tipa}{T3}{cmr}{m}{n}
\DeclareMathAccent{\invbreve}{\mathalpha}{tipa}{16}

\renewcommand{\ker}{\operatorname{ker}}

\newcommand{\ran}{\operatorname{ran}}

\newcommand{\MSE}{\operatorname{MSE}}

\newcommand{\dx}{\operatorname{d}\!x}
\newcommand{\dhx}{\operatorname{d}\!\hat{x}}
\newcommand{\dy}{\operatorname{d}\!y}
\newcommand{\dz}{\operatorname{d}\!z}
\newcommand{\PG}{\operatorname{PG}}
\newcommand{\Tr}{\operatorname{Tr}}
\newcommand{\Par}[1]{\left( #1 \right)}

\definecolor{champagne}{rgb}{0.97, 0.91, 0.81}
\newcommand{\Frame}[1]{\begin{tcolorbox}[colframe = white, colback= blue!8!white, breakable, enhanced]#1 \end{tcolorbox}}

\theoremstyle{oupplain}
\newtheorem{theorem}{Theorem}[section]
\newtheorem{lemma}[theorem]{Lemma}
\newtheorem{proposition}[theorem]{Proposition}
\newtheorem{corollary}[theorem]{Corollary}
\theoremstyle{oupdefinition}
\newtheorem{definition}{Definition}[section]
\theoremstyle{oupremark}
\newtheorem{remark}[theorem]{Remark}

\theoremstyle{oupproof}

\definecolor{ashgrey}{rgb}{0.7, 0.75, 0.71}
\newenvironment{proof}{
	\tcolorbox[blanker,breakable,left=5mm,parbox=false,
    before upper={\parindent15pt},
    after skip=10pt,
	borderline west={0.7mm}{0pt}{blue!8!white}]
    {\noindent{\it \textbf{Proof:}}}
}{
    \textcolor{black}{\hbox{}\nobreak\hfill$\blacksquare$} 
    \endtcolorbox
}

\numberwithin{equation}{section}

\articletype{RESEARCH ARTICLE}

\begin{document}
\setlength{\abovedisplayskip}{12pt}
\setlength{\belowdisplayskip}{12pt}

\begin{Frontmatter}
\title{Near-optimal performance of square-root measurement for general score functions and quantum ensembles}

    \author{Hemant K. Mishra \hspace{-0.10cm}\thanks{Department of Mathematics and Computing, Indian Institute of Technology (ISM) Dhanbad, Jharkhand 826004, India; {\email{hemantmishra1124@iitism.ac.in}}} \hspace{-0.15cm}\thanks{School of Electrical and Computer Engineering, Cornell University, Ithaca, New York~14850, USA; {\email{hemant.mishra@cornell.edu}}}, Ludovico Lami \hspace{-0.15cm} \thanks{Scuola Normale Superiore, Piazza dei Cavalieri 7, 56126 Pisa, Italy; \email{ludovico.lami@gmail.com}}, and 
    Mark M. Wilde \hspace{-0.15cm} \thanks{School of Electrical and Computer Engineering, Cornell University, Ithaca, New York~14850, USA; \email{wilde@cornell.edu}}
    }

    \abstract
    {
        The Barnum–Knill theorem states that the optimal success probability in the multiple state
    discrimination task is not more than the square root of the success probability when the pretty good  or square-root measurement is used for this task. An assumption of the theorem is that the underlying ensemble consists of finitely many quantum states over a finite-dimensional quantum system.
    Motivated in part by the fact that the success probability is not a relevant metric for continuous ensembles, in this paper we provide a generalization of the notion of pretty good measurement and the Barnum--Knill theorem for general quantum ensembles, including those described by a continuous parameter space and an infinite-dimensional Hilbert space. 
    To achieve this, we also design a general metric of performance for quantum measurements that generalizes the success probability, namely, the \emph{expected gain} of the measurement with respect to a \emph{positive score function}. 
    A notable consequence of the main result is that, in a Bayesian estimation task, the mean square error of the generalized pretty good measurement does not exceed twice the optimal mean square error. 
    }
\end{Frontmatter}

\Frame{\tableofcontents}

\section{Introduction}

    Quantum estimation theory is crucial for understanding the extraction of accurate information from a quantum system, and it is central in metrology, quantum sensing, and quantum information processing.
    The framework involves a parameter space $\mathscr{X}$ and a quantum system such that each parameter value $x \in \mathscr{X}$ corresponds to a state $\rho_x$ of the quantum system.
    The aim is to estimate the parameter $x$ by performing a quantum measurement on the system.
    However, the very nature of quantum mechanics precludes exhaustive measurements of the system.
    The quantum estimation problem thus requires designing a measurement procedure that involves a quantum measurement scheme and a quantity to be measured about the system. 
    Mathematically, it takes the form of an optimization problem over \emph{probability operator-valued measures} (POVMs)\footnote{The acronym POVM generally stands for \emph{positive} operator-valued measure in the literature. We call it \emph{probability} operator-valued measure to mean the same.
    This helps us distinguish POVMs from general countably additive operator-valued maps that do not necessarily add to the identity map. This distinction is useful to develop the integration theory with respect to countably additive operator-valued maps and observe some properties needed in the proof of Theorem~\ref{thm:generalized_bk}. See Appendix~\ref{app:int_wrt_povm}.}, where the objective function is the quantity to be measured.
    A measurement scheme achieving the optimal measurement value is called an optimal measurement.
    Depending on the choice of the objective function to be optimized, particular examples of quantum estimation problems include quantum state discrimination~\cite{helstrom1969quantum, Holevo1976} (see also~\cite{Barnett09}) 
    and quantum state exclusion~\cite{bandyopadhyay_etal2014, Heinosaari_2018, Russo_etal2023, mishra2024optimal} 
    for the case when the parameter space $\mathscr{X}$ is finite. 

\subsection{Literature review}

    It is challenging in general to determine an explicit optimal measurement scheme when a quantum system is prepared in a state selected from a non-orthogonal set of quantum states, and it does not seem possible to explicitly specify it even in the simplest case in which the parameter space is finite with more than two elements. 
    An optimal measurement scheme for discriminating two quantum states with the maximum possible success probability is explicitly given by the Helstrom{--Holevo} measurement~\cite{helstrom1969quantum,Holevo1976}.
    Although there is no explicit form of an optimal quantum measurement known for the state discrimination task if the parameter space is finite with more than two elements, a measurement called the \emph{pretty good measurement} or \emph{square-root measurement}~\cite{Belavkin75, Belavkin75a, H79, hausladen1994pretty} is known to perform ``pretty well''.
    
    The pretty good measurement (PGM) can be constructed canonically from the given ensemble of quantum states.
    Concretely, let $\rho_1,\ldots, \rho_r$ be quantum states of a finite-dimensional quantum system, and let $p_1,\ldots, p_r$ be a prior probability distribution of the parameter values $1,\ldots, r$ with full support.
    Consider a quantum system prepared in a state selected randomly from the ensemble $\mathscr{E}\coloneqq((p_i, \rho_i))_{i=1}^r$.
    Let $\rho \coloneqq \sum_{i=1}^r p_i \rho_i$ be the average state of the ensemble.
    For a given POVM $M\coloneqq(M_1,\ldots, M_r)$ and as a consequence of the Born rule, the average success probability for identifying the true state of the system is as follows:
    \begin{align}
        P_{s}(\mathscr{E}, M) \coloneqq \sum_{i=1}^r p_i \Tr \! \left[M_i \rho_i  \right].
    \end{align}
    As such, the mathematical optimization problem corresponding to state discrimination is given by
    \begin{align}
        \sup_{M \in \operatorname{POVM}} P_{s}(\mathscr{E}, M).  
    \end{align}
    The PGM is defined as the POVM $M^{\PG} \equiv (M^{\PG}_1, \ldots, M^{\PG}_r)${, where}
    \begin{align}\label{eq:pgm}
        M_i^{\PG} \coloneqq \sqrt{\rho^+} \, p_i \rho_i \, \sqrt{\rho^+} + p_i \Pi_{\ker(\rho)}
    \end{align}
    for $1 \leq i \leq r$.
    Here $\rho^+$ denotes the Moore--Penrose inverse of $\rho$ and $\Pi_{\ker (\rho)}$ denotes the orthogonal projection onto the kernel of $\rho$.
    The Barnum--Knill theorem
    states that the PGM is ``pretty good'' for the quantum state discrimination task, in the sense that the optimal success probability 
    does not exceed the square root of the success probability when using the PGM \cite[Corollary~3]{BarnumKnill}; i.e.,
        \begin{align} \label{eq:bk}
            \sup_{M \in \operatorname{POVM}} P_s(\mathscr{E},M) \leq \sqrt{P_s(\mathscr{E},M^{\PG})}.
        \end{align}
    The PGM has also been shown to be an optimal measurement scheme for the state discrimination task for some specific types of quantum ensembles (see, e.g.,~\cite{eldar2001, eldar2003, bacon2005optimal, mochon2006, leditzky2022optimality}).

    Generalizing the Barnum--Knill theorem in~\eqref{eq:bk} to continuous ensembles poses a challenge, as the success probability does not appear to be a meaningful figure of merit in this case --- since the space of measurement outcomes is continuous, the success probability is exactly equal to zero in most cases of interest. 
    In addition, we also aim to extend the theorem to infinite-dimensional Hilbert spaces, which likewise falls outside the scope of its original formulations. 
    
    Both of these technical challenges arise naturally in the context of bosonic Gaussian ensembles, a setting of  practical relevance. In that framework, the parameter space is $\mathds{R}^{2n}$, and $\rho_x$ is an $n$-mode bosonic Gaussian state obtained by displacing a fixed Gaussian state $\rho_0$ by the Weyl displacement operator parametrized by $x \in \mathds{R}^{2n}$~\cite{holevo2020gaussian, holevo2021classical, holevo_accessible_2021, mlmw2024}. The PGM has been constructed for these ensembles, and some appealing properties have been established: to wit,
    the measurement is Gaussian and achieves the accessible information of the corresponding ensemble~\cite{holevo2020gaussian}.
    Also, an explicit form of the mean square error of the PGM has been determined in~\cite[Theorem~2]{mlmw2024} for a Bayesian estimation task.
    However, to the best of our knowledge, no analysis of the performance of the PGM for bosonic Gaussian ensembles exists in the literature; also, there is no evidence of near-optimality of these measurements in the parameter estimation task, similar to what is known for the case of a finite parameter space. 
    One of our results (Theorem~\ref{thm:mse}) indeed proves near-optimality of such measurements in the Bayesian estimation of unknown parameters.
    This, combined with the fact that these measurements are Gaussian, makes the PGM a desirable theoretical and practical tool in the quantum estimation task for bosonic Gaussian ensembles.

\subsection{Summary of contributions}

    In this paper, we provide a complete generalization of the notion of PGM and the Barnum--Knill theorem for general quantum ensembles, including ensembles with a continuous parameter space and infinite-dimensional underlying quantum system.
    As an essential ingredient of our setting, we consider a general metric of performance of quantum measurements that suitably generalizes the success probability, namely,
    the \emph{expected gain} of the measurement with respect to a \emph{positive score function}. 
    As an application of our findings, we show that the optimal mean square error in a Bayesian estimation task
    is bounded from below by half of the mean square error of the generalized PGM. 
    This also justifies that this measurement is ``pretty good'' for the quantum estimation task.	
        
    The following is a detailed description of our main results:
    \begin{enumerate}
        \item We extend the notion of PGM to general quantum ensembles by canonically constructing the \emph{generalized pretty good measurement} (GPGM), and  Theorem~\ref{thm:pgm} establishes that the GPGM is a legitimate quantum measurement.
                
        \item Theorem~\ref{thm:generalized_bk} establishes a generalization of the Barnum--Knill theorem with respect to the GPGM for general quantum ensembles.
        In a nutshell, we prove that, for a given ensemble and positive score function, the optimal expected gain 
        does not exceed the square root of the expected gain of the GPGM.
            
        \item As an application of the above findings in the quantum estimation problem,   Theorem~\ref{thm:mse} establishes that the mean square error of the GPGM in a Bayesian estimation task
        is not more than twice  the optimal mean square error.
        
        \item We state and prove two mathematically interesting results related to integration of real-valued functions with respect to operator-valued measures.
        These results are needed for rigorously proving the main results of our paper.
    	\begin{enumerate}
                \item Theorem~\ref{thm:int_trace_class_valued_measure} states that the integration of a bounded real-valued measurable function with respect to a self-adjoint positive trace-class operator-valued measure is a self-adjoint trace-class operator. 
                
                \item Theorem~\ref{thm:int_hilbert_schmidt_valued_measure} states that the integration of a bounded real-valued measurable function with respect to a self-adjoint positive Hilbert--Schmidt operator-valued measure is a self-adjoint Hilbert--Schmidt operator. 
    	\end{enumerate}
    \end{enumerate}
        
\subsection{Paper organization}

    The rest of our paper is organized as follows. We set 
    the notation and review basic mathematical concepts relevant to our work in Section~\ref{sec:back}.
    The main results of our paper are presented in Section~\ref{sec:main}: the construction of the GPGM and its legitimacy as a measurement are given in Section~\ref{sec:gpgm}, the generalized Barnum--Knill theorem is established in Section~\ref{sec:gbk}, and an analysis of the mean square error of the GPGM in Bayesian estimation is provided in Section~\ref{sec:mse}.
    Appendices~\ref{app:int_wrt_povm} through \ref{app:exp_gain_integral} contain rigorous mathematical analyses of the concepts required for proving the main results.
    
\section{Background and notation}

    \label{sec:back}
       
    Let $\mathscr{H}$ be a complex separable Hilbert space. In what follows, we adopt the following notation to indicate various sets of operators acting on $\mathscr{H}$.
    \begin{itemize}
        \item $\mathscr{L}(\mathscr{H})$: Banach space of bounded linear operators, equipped with the operator norm~$\left\|\cdot \right\|$.
        \item $\mathscr{L}_t(\mathscr{H})$: separable Banach space of all trace-class operators, equipped with the trace norm $\left\|\cdot \right\|_1$.
        \item $\mathscr{L}_{st}(\mathscr{H})$: real space of self-adjoint trace-class operators.
        \item $\mathscr{D}(\mathscr{H})$: set of \emph{density operators}, i.e., self-adjoint positive trace-class operators with trace equal to one. 
        \item $\mathscr{L}_{\operatorname{HS}}(\mathscr{H})$: Hilbert space of Hilbert--Schmidt operators equipped with the Hilbert--Schmidt norm $\left\| \cdot \right\|_{2}$. 
        \item $\mathscr{L}_{s\operatorname{HS}}(\mathscr{H})$: real Hilbert space of self-adjoint Hilbert--Schmidt operators. 
    \end{itemize}
    For each of these sets, we might want to consider the positive semi-definite component only; we do so by appending a subscript `$+$': for example, let $\mathscr{L}_s(\mathscr{H})_+$ denote the set of bounded (self-adjoint) positive semi-definite operators acting on $\mathscr{H}$. Note that every $\varrho \in \mathscr{L}_s(\mathscr{H})_+$ admits a unique square root $\varrho^{\frac{1}{2}} \in \mathscr{L}_s(\mathscr{H})_+$~\cite[Section~104]{Riesz_Nagy_1957}.
    
    Given $\varrho \in \mathscr{L}(\mathscr{H})$, let $\ker(\varrho)$ denote  the kernel of $\varrho$, and let $\ran(\varrho)$ denote the range of $\varrho$.

    A generalized quantum ensemble $\mathscr{E} \coloneqq \left(\left(\mu(\dx), \rho_x \right)\right)_{x \in \mathscr{X}}$ consists of a probability measure~$\mu$ on a locally compact Hausdorff space $\mathscr{X}$ and a Borel measurable function $\mathscr{X} \ni x \mapsto \rho_x \in \mathscr{D}(\mathscr{H})$.
    The average state of the ensemble is defined as the barycenter of $\mu$, given by the \emph{Bochner integral} (see Appendix~\ref{app:bochner_integral}):
    \begin{align}\label{eq:average_state_integral}
        \rho \coloneqq \int_{\mathscr{X}}  \rho_x  \, \mu(\dx).
    \end{align}
    
    Let $\mathscr{B}(\mathscr{X})$ denote the Borel $\sigma$-algebra of $\mathscr{X}$.
    A map $m\colon \mathscr{B}(\mathscr{X}) \to \mathscr{L}_s(\mathscr{H})$ is called an operator-valued measure if it is countably additive in the strong operator topology; i.e., $m\!\left(\bigcup_{n \in \mathds{N}} E_n \right)=\sum_{n \in \mathds{N}} m(E_n)$
    for every sequence $\left(E_n \right)_{n \in \mathds{N}}$ of disjoint sets in $\mathscr{B}(\mathscr{X})$, and the convergence of the series is in the sense of the strong operator topology.\footnote{{A sequence of bounded operators $(X_n)_{n\in \mathds{N}}$ is said to converge to a bounded operator $X$ in the strong operator topology if, for every fixed vector $\ket{\psi}\in \mathscr{H}$, it holds that $\lim_{n\to\infty} \left\| X_n\ket{\psi} - X \ket{\psi} \right\| = 0$.}}
    An operator-valued measure $m$ is called a probability operator-valued measure (POVM) if $m(E) \in \mathscr{L}_s(\mathscr{H})_+$ for all $E \in \mathscr{B}(\mathscr{X})$ and $m(\mathscr{X})$ is equal to the identity operator acting on $\mathscr{H}$.
    A general quantum measurement of a quantum system with the underlying Hilbert space $\mathscr{H}$ and the outcome space~$\mathscr{X}$ is mathematically described by a POVM $m\colon \mathscr{B}(\mathscr{X}) \to \mathscr{L}_s(\mathscr{H})_+$. 
    Let $\mathscr{M}(\mathscr{X}, \mathscr{L}_s(\mathscr{H})_+)$ denote the set of all POVMs.

    We refer the reader to~\cite{hs_2006, shirokov2024average, HOLEVO1973337} for mathematical details 
    on the general theory of quantum ensembles and measurements (see also~\cite{holevo_accessible_2021}).

\section{Main results}

    \label{sec:main}
    
\subsection{Generalized pretty good measurement}

    \label{sec:gpgm}
    
    Let $\mathscr{E} \coloneqq \left(\left(\mu(\dx), \rho_x\right)\right)_{x \in \mathscr{X}}$ be a quantum ensemble with underlying complex separable Hilbert space $\mathscr{H}$.
    Let $\rho$ be the average state of the ensemble given by the Bochner integral~\eqref{eq:average_state_integral}.
    For $E \in \mathscr{B}(\mathscr{X})$, set 
    \begin{align}
        \rho_{E} \coloneqq \int_{E}  \rho_{x} \, \mu(\dx).
    \end{align}
    Since $\rho_E \leq \rho$, we can apply~\cite[p.~249, Ex.~10.8.8]{schmudgen} to conclude that there exists a unique operator $\Lambda_E \in \mathscr{L}(\mathscr{H})$ satisfying 
    \begin{equation} \label{eq:Lambda_E}
        \|\Lambda_E \| \leq 1,\qquad \operatorname{ker}(\rho) \subseteq \operatorname{ker}(\Lambda_E),\qquad \Lambda_E \rho^{\frac{1}{2}} = \rho_{E}^{\frac{1}{2}}.
    \end{equation}
    \Frame{
    \begin{definition}[Generalized pretty good measurement]
    \label{def:pgm}
    Given a quantum ensemble $\mathscr{E} \coloneqq \left(\left(\mu(\dx), \rho_x\right)\right)_{x \in \mathscr{X}}$ on a separable Hilbert space $\mathscr{H}$, the associated \emph{generalized pretty good measurement} (GPGM) is the map
    $m^{\PG}\colon \mathscr{B}(\mathscr{X}) \to \mathscr{L}_s(\mathscr{H})_+$ defined by
    \begin{align}
        m^{\PG}(E)\coloneqq \Lambda_E^{\dagger} \Lambda_E^{\vphantom{\dag}} + \mu(E) \Pi_{\operatorname{ker}(\rho)},
    \end{align}
    for all $E \in \mathscr{B}(\mathscr{X})$. Here, $\Lambda_E$ is given by~\eqref{eq:Lambda_E}, and $\Pi_{\ker(\rho)}$ is the orthogonal projection onto the kernel of $\rho$. 
    \end{definition}}
    
    It is worth observing that $\rho^{\frac{1}{2}} m^{\PG}(E) \rho^{\frac{1}{2}} = \rho_{E}$ and $\operatorname{ran}\big(\Lambda_E^{\dagger} \Lambda_E^{\vphantom{\dag}}\big) \subseteq \operatorname{ran}( \rho)$.

    \Frame{
    \begin{theorem}\label{thm:pgm}
     Let $\mathscr{E} \coloneqq \left(\left(\mu(\dx), \rho_x\right)\right)_{x \in \mathscr{X}}$ be a quantum ensemble with underlying complex separable Hilbert space $\mathscr{H}$. The associated generalized pretty good measurement, constructed in Definition~\ref{def:pgm}, is a POVM, i.e., $m^{\PG} \in \mathscr{M}(\mathscr{X}, \mathscr{L}_s(\mathscr{H})_+)$.
    \end{theorem}}

    \begin{proof}
    By definition, we have that  $m^{\PG}(E) \in \mathscr{L}_s(\mathscr{H})_+$ for all $E \in \mathscr{B}(\mathscr{X})$.
    Also, {by direct inspection of~\eqref{eq:Lambda_E} we see that} $\Lambda_{\mathscr{X}}=\Pi_{\ker(\rho)^{\perp}}$ is the orthogonal projection onto the orthogonal complement of the kernel of $\rho$ {(i.e., the support of $\rho$)}.
    This implies that $m^{\PG}(\mathscr{X})=\Pi_{\ker(\rho)^{\perp}}+\Pi_{\ker(\rho)}$, which is the identity operator on $\mathscr{H}$.
    It now remains to establish countable additivity of $m^{\PG}$.
    
    Let $\left(E_n \right)_{n\in \mathds{N}}$
    be a sequence of disjoint sets in $\mathscr{B}(\mathscr{X})$.
    By applying the spectral theorem for self-adjoint positive compact operators, there exists an orthonormal basis $\left\{|e_n\rangle: n \in \mathds{N} \right\}$ of $\mathscr{H}$ and a sequence $\left(q_n \right)_{n\in \mathds{N}}$ of nonnegative numbers, such that $\sum_{n \in \mathds{N}} q_n < \infty$ and $\rho |e_n \rangle = q_n |e_n \rangle$ for all $n \in \mathds{N}$.
    Set $\mathds{N}_{\rho>0} \coloneqq \{n \in \mathds{N}: q_n > 0\}$.
    The condition $\operatorname{ker}(\rho) \subseteq \operatorname{ker}(\Lambda_E)$ for all $E \in \mathscr{B}(\mathscr{X})$ implies that, for $k \in \mathds{N}\backslash \mathds{N}_{\rho>0}$, 
    \begin{align}
        m^{\PG}\!\left(\bigcup\nolimits_{n \in \mathds{N}} E_n\right)|e_k\rangle
            &= \mu\left(\bigcup\nolimits_{n \in \mathds{N}} E_n\right)|e_k\rangle \\
            &= \sum_{n \in \mathds{N}} \mu(E_n)|e_k\rangle \\
            &= \sum_{n \in \mathds{N}} m^{\PG}(E_n)|e_k\rangle.
    \end{align}
    We have thus shown that the operators $m^{\PG}\big(\bigcup_{n} E_n\big)$ and $\sum_{n} m^{\PG}(E_n)$ agree on $\operatorname{ker}(\rho)$.
    We now argue that these operators also coincide on 
    \begin{equation}
    \operatorname{ker}(\rho)^{\perp}=\overline{\operatorname{span} \{|e_k\rangle: k \in \mathds{N}_{\rho>0}\}}.
    \end{equation}
    Let $k \in \mathds{N}_{\rho>0}$ be arbitrary.
    Observe that, for all $E \in \mathscr{B}(\mathscr{X})$,
    \begin{align}
    \overline{\operatorname{ran}\!\big(\Lambda_E^\dagger \Lambda_E^{\vphantom{\dag}}\big)}= \ker(\Lambda_E)^\perp 
    \subseteq \ker(\rho)^\perp.
    \end{align}
    Therefore, we find that
    \begin{equation}
    m^{\PG}\!\left(\bigcup\nolimits_{n \in \mathds{N}} E_n  \right) |e_k \rangle \in \ker(\rho)^\perp, \quad  \sum_{n \in \mathds{N}} m^{\PG}\!\left(E_n\right) |e_k \rangle \in \ker(\rho)^\perp.
    \end{equation}
    For all $k^{\prime} \in \mathds{N}_{\rho>0}$, consider that
    \begin{align}
         \left\langle e_{k^{\prime}}\right|m^{\PG}\!\left(\bigcup\nolimits_{n \in \mathds{N}} E_n\right) \left|e_k \right\rangle 
            &= \dfrac{1}{\sqrt{q_{k^{\prime}}q_k}} \left\langle e_{k^{\prime}}\right| \rho^{\frac{1}{2}}
            m^{\PG}\!\left(\bigcup\nolimits_{n \in \mathds{N}} E_n\right) \rho^{\frac{1}{2}} \left|e_k \right\rangle \\
            &= \dfrac{1}{\sqrt{q_{k^{\prime}}q_k}} \langle e_{k^{\prime}}|\rho_{\bigcup_n E_n} |e_k \rangle \\
            &= \dfrac{1}{\sqrt{q_{k^{\prime}}q_k}} \left\langle e_{k^{\prime}}\right|\int_{\bigcup_n E_n}  \rho_{x} \mu(\dx) \left|e_k \right \rangle \\
            &=\dfrac{1}{\sqrt{q_{k^{\prime}}q_k}} \left\langle e_{k^{\prime}}\right|\sum_{n \in \mathds{N}}\int_{E_n}  \rho_{x} \mu(\dx) \left|e_k \right \rangle \\
            &= \dfrac{1}{\sqrt{q_{k^{\prime}}q_k}} \left\langle e_{k^{\prime}}\right|\sum_{n \in \mathds{N}} \rho_{E_n} \left|e_k \right \rangle \\
            &= \dfrac{1}{\sqrt{q_{k^{\prime}}q_k}} \left\langle e_{k^{\prime}}\right| \rho^{\frac{1}{2}} \left(\sum\nolimits_{n \in \mathds{N}}  m^{\PG}(E_n) \right) \rho^{\frac{1}{2}} \left|e_k \right \rangle \\
            &= \left\langle e_{k^{\prime}}\right|\sum_{n \in \mathds{N}} m^{\PG}\!\left(E_n\right) \left|e_k \right\rangle.
    \end{align}
    This implies that the operators $m^{\PG}(\bigcup_{n} E_n)$ and $\sum_{n} m^{\PG}(E_n)$ also coincide on $\operatorname{ker}(\rho)^\perp$.
    \end{proof}
        
    Let us emphasize that the known PGM is a special case of the GPGM.
    Indeed, consider the set $\mathscr{X} \coloneqq \{1,\ldots, r\}$ with the discrete metric on it and a probability distribution $p_1,\ldots, p_r$ in Definition~\ref{def:pgm}.
    We know that $\sqrt{\rho^+} \sqrt{\rho}$ is the orthogonal projection onto the range of $\sqrt{\rho}$.
    Since $\ran(\sqrt{p_i \rho_i}) \subseteq \ran(\sqrt{\rho})$, we thus get 
    \begin{align}
        \sqrt{p_i \rho_i} \sqrt{\rho^+} \sqrt{\rho}= \sqrt{p_i \rho_i},
    \end{align}
    so that $\Lambda_{\{i\}} = \sqrt{p_i \rho_i} \sqrt{\rho^+}$.
    By definition, we thus obtain the POVM
    \begin{align}
         \Lambda_{\{i\}}^\dagger \Lambda_{\{i\}}^{\vphantom{\dag}} = \sqrt{\rho^+} p_i \rho_i \sqrt{\rho^+}+ p_i \Pi_{\ker(\rho)}
    \end{align}
    for $1 \leq i \leq r$, which coincides with the definition of the PGM given by~\eqref{eq:pgm}.   

\subsection{Generalized Barnum--Knill theorem}

    \label{sec:gbk}

The fundamental object of interest in the setting of the Barnum--Knill theorem is the success probability. As mentioned in the introduction, for a generic continuous ensemble, this is not necessarily a relevant metric --- since there are continuously many possible outcomes labeled by $x$, the probability of guessing the correct one \emph{exactly} is strictly equal to zero. In practice, however, we might be satisfied with an \emph{approximately} correct guess. To formalize this idea, we introduce the notion of a \emph{score function}, whose value on a pair $(x,\hat{x})$ quantifies, with a real number between $0$ and $1$, the quality of the guess $\hat{x}$ when the real value of the variable is $x$. Our result  applies to a specific subset of score functions, namely, the positive ones, which we now define formally.

\subsubsection{Positive score functions}
    Suppose $\mathscr{Y}$ and $\mathscr{Z}$ are measurable spaces.
    Let $S, T\colon \mathscr{Y} \times \mathscr{Z} \to \mathds{R}$ be measurable functions.
    For a measure $\pi$ on $\mathscr{Z}$, we define a \emph{convolution} $S \star_\pi T \colon \mathscr{Y} \times \mathscr{Y} \to \mathds{R}$ by
    \begin{align}
            (S \star_\pi T)(x, y) \coloneqq \int_{\mathscr{Z}}  S(x, z) T(y, z) \pi(\dz), \qquad \forall x, y \in \mathscr{Y},
    \end{align}
    whenever the above integral exists for all $(x, y) \in \mathscr{Y}\times \mathscr{Y}$.

    \Frame{
    \begin{definition}\label{def:score_func}
        Suppose $\mathscr{X}$ is a measurable space.
        We call a measurable function $S\colon \mathscr{X} \times \mathscr{X} \to [0,1]$ a \emph{positive score function} on $\mathscr{X}$ if there exists another measurable space $\mathscr{Y}$, a measure $\pi$ on $\mathscr{Y}$, and a measurable function $P\colon \mathscr{X} \times \mathscr{Y} \to \mathds{R}$ such that
        \begin{enumerate}
            \item $P$ is bounded,
            \item the convolution $P \star_{\pi} P$ is well defined, and $S=P\star_{\pi} P$.
        \end{enumerate}
    \end{definition}
    }

    In the case of a finite alphabet $\mathscr{X}$, which we can identify with $\{1,\ldots, r\}$, positive score functions labeled by 
    $S$ are in one-to-one correspondence with positive semi-definite matrices, through the map $S \mapsto \big(S(x,\hat{x})\big)_{x,\hat{x}=1}^r$.
    Some other obvious examples of positive score functions include constant functions: for every $a \in [0,1]$, define $C_a(x,y)=a$ for all $x,y \in \mathscr{X}$; one can then see that for every Borel probability measure $\pi$ on $\mathscr{X}$, we have $C_a = C_{\sqrt{a}} \star_{\pi} C_{\sqrt{a}}$.
    
    Another class of examples involves Gaussian functions. For an $N \times N$ real symmetric positive definite matrix $\Sigma$, let $g_{\Sigma}$ denote the Gaussian distribution on $\mathds{R}^{N}$ given by
     \begin{align}
       g_{\Sigma}(x) \coloneqq  \dfrac{1}{(2\pi)^{N/2}\sqrt{\det \Sigma}} \exp\!\left[-\dfrac{1}{2} x^\intercal \Sigma^{-1}x\right], \qquad \forall x \in \mathds{R}^N.
    \end{align}
    \Frame{
    \begin{definition}
        \label{def:gaussian_score_function}
        Let $\Sigma$ be an $N \times N$ real symmetric positive definite matrix. 
    Define $S_\Sigma\colon  \mathds{R}^{N} \times \mathds{R}^{N} \to [0,1]$ by
	\begin{align}
        \label{eq:gaussian_score_function}
		S_\Sigma(x,y) \coloneqq (2\pi)^{N/2}\sqrt{\det \Sigma} \ g_{\Sigma}(x-y), \qquad x, y \in \mathds{R}^{N}.
	\end{align}
    It follows by Proposition~\ref{prop:gaussian_score_convolution} in Appendix~\ref{app:gaussian_score} that $S_\Sigma$ is a positive score function, and we call it a \emph{Gaussian score function}.
    \end{definition}
    }
    
\subsubsection{Setup for estimation problem}

    In the quantum estimation problem, the state of a quantum system is selected from an ensemble $\mathscr{E}\coloneqq\left(\left(\mu(\dx), \rho_x\right)\right)_{x \in \mathscr{X}}$, and a decision strategy is specified by a POVM $m \in \mathscr{M}(\mathscr{X}, \mathscr{L}_s(\mathscr{H})_+)$.
    The measurement $m$ yields an estimate of an unknown parameter; for a given true parameter $x$ and a measurable set $E \in \mathscr{B}(\mathscr{X})$, the probability that the estimated value $\hat{x}$ lies in $E$ is given by
    \begin{align}
        \label{eq:cond_probability_POVM}
        \mathds{P}\!\left(\hat{x} \in E \vert x \right) \coloneqq 
        \Tr \!\left[\rho_x m(E)\right].
    \end{align}
    We now assume that there is a positive score function $S(x, \hat{x})$ that specifies a relative positive score of an estimate $\hat{x}$ of the parameter $x$.
    The expected gain function of the parameter $x$ for the POVM is given by 
    \begin{align}
        G_{\mathscr{E}, S, m}(x) \coloneqq \Tr \!\left[\rho_x \int_{\mathscr{X}} S(x,\hat{x}) m(\dhx) \right]. \label{eq:gain_function}
    \end{align}
    Since the map $\hat{x} \mapsto S(x, \hat{x})$ is {non-negative, bounded, and} measurable, which follows from~\cite[Proposition~1.2.24]{hytonen2016analysis}, {and moreover $m$ is a POVM,} the integral in~\eqref{eq:gain_function} is well defined, and we have $0 \leq G_{\mathscr{E}, S, m}(x) \leq 1$ (see Appendix~\ref{app:int_wrt_povm}).
    Furthermore, the map $G_{\mathscr{E}, S, m}: \mathscr{X} \to [0, 1]$ is measurable (see Appendix~\ref{app:exp_gain_integral}).

    The expected gain of the POVM on the ensemble is then given by
    \begin{align}
        G_{\mathscr{E}, S, m} \coloneqq \int_{\mathscr{X}} G_{\mathscr{E}, S, m}(x) \mu(\dx). \label{eq:exp_gain}
    \end{align}
    The quantum estimation problem consists of finding a POVM that maximizes the expected gain~\eqref{eq:exp_gain}; i.e.,
    the aim is to find a solution to the following optimization problem:
    \begin{align}
        G_{\mathscr{E}, S} \coloneqq \sup_{m \in \mathscr{M}(\mathscr{X}, \mathscr{L}_s(\mathscr{H})_+)} \int_{\mathscr{X}} \Tr \!\left[\rho_x \int_{\mathscr{X}} S(x,\hat{x}) m(\dhx) \right] \mu(\dx).
    \end{align}
    We call $G_{\mathscr{E}, S}$ the optimal expected gain for the ensemble $\mathscr{E}$ and a given positive score function~$S$.

\subsubsection{Near-optimal gain of generalized pretty good measurement}
    Our next result states that the GPGM achieves a near-optimal expected gain.
    This is in the sense that the square root of the expected gain of the GPGM is an upper bound on the optimal expected gain.
        
        \Frame{
        \begin{theorem}
        \label{thm:generalized_bk}
           Let $\mathscr{E}\coloneqq \left(\left(\mu(\dx), \rho_x\right)\right)_{x \in \mathscr{X}}$ be a quantum ensemble, and let $S:\mathscr{X} \times \mathscr{X} \to [0, 1]$ be a positive score function, as given in Definition~\ref{def:score_func}.
           Then the optimal expected gain for the ensemble and the given score function does not exceed the square root of the expected gain of the generalized pretty good measurement; i.e., the following inequality holds:
           \begin{align}
               G_{\mathscr{E}, S} \leq \sqrt{G_{\mathscr{E}, S, m^{\PG}}},
           \end{align}
           {where $m^{\PG}$ is the associated generalized pretty good measurement (see Definition~\ref{def:pgm}).}
        \end{theorem}
        }
        
    \begin{proof}
        Let $m \in \mathscr{M}(\mathscr{X}, \mathscr{L}_s(\mathscr{H})_+)$ be arbitrary.
        From~\eqref{eq:gain_function} and \eqref{eq:exp_gain}, we have that
        \begin{align}
            \label{eq:exp_gain_calc_1}
            G_{\mathscr{E}, S, m}
                &= \int_{\mathscr{X}} \Tr \!\left[\rho_x \int_{\mathscr{X}} S(x,\hat{x}) m(\dhx) \right] \mu(\dx).
        \end{align}
        By using the cyclicity of the trace function and considering the operator-valued measure $\rho_x^{\frac{1}{2}} m \left(\cdot \right)\rho_x^{\frac{1}{2}}$ in Theorem~\ref{thm:int_trace_class_valued_measure}, it follows that the trace function in \eqref{eq:exp_gain_calc_1} can be taken inside the integral, and we conclude that
        \begin{align}
            \label{eq:exp_gain_trace_inside_integral}
            G_{\mathscr{E}, S, m}
                &= \int_{\mathscr{X}}  \int_{\mathscr{X}} S(x,\hat{x}) \Tr \!\left[\rho_x m(\dhx) \right] \mu(\dx).
        \end{align}
        Since $S$ is a positive score function, according to Definition~\ref{def:score_func}, there exists a measurable space $\mathscr{Y}$, a measure $\pi$ on $\mathscr{Y}$, and a bounded measurable map $P: \mathscr{X} \times \mathscr{Y} \to \mathds{R}$ such that $S=P \star_\pi P$.
        We then conclude from \eqref{eq:exp_gain_trace_inside_integral} that
        \begin{align}
            G_{\mathscr{E}, S, m}
                &= \int_{\mathscr{X}}  \int_{\mathscr{X}} \left(\int_{\mathscr{Y}} P(x, y)P(\hat{x}, y) \pi(\dy) \right)  \Tr \!\left[\rho_x m(\dhx) \right] \mu(\dx).
                \label{eq:proof-mid-step-1}
        \end{align}
        By Fubini's theorem~\cite[Theorem~5.2.2]{cohn2013measure}, the order of integration in \eqref{eq:proof-mid-step-1} can be interchanged, leading to
        \begin{align}
            G_{\mathscr{E}, S, m}
                &= \int_{\mathscr{Y}}  \left(\int_{\mathscr{X}} \int_{\mathscr{X}} P(x, y)P(\hat{x}, y)  \Tr \!\left[\rho_x m(\dhx) \right] \mu(\dx)  \right) \pi(\dy).
        \end{align}
        Again, by using the cyclicity of the trace function and by considering the operator-valued measure $\rho_x^{\frac{1}{2}} m \left(\cdot \right)\rho_x^{\frac{1}{2}}$ in Theorem~\ref{thm:int_trace_class_valued_measure}, we conclude that 
        \begin{align}
            G_{\mathscr{E}, S, m}
                &= \int_{\mathscr{Y}}  \int_{\mathscr{X}} \Tr \!\left[ \int_{\mathscr{X}} P(x, y)P(\hat{x}, y)  \rho_x m(\dhx) \right] \mu(\dx)  \pi(\dy).
        \end{align}
        Since the trace map is bounded on the space of trace-class operators, we conclude from~\eqref{eq:bochner_int_bdd_map_action} in Appendix~\ref{app:bochner_integral} that
        \begin{align}
            G_{\mathscr{E}, S, m}
                &= \int_{\mathscr{Y}} \Tr \!\left[ \int_{\mathscr{X}}  \int_{\mathscr{X}} P(x, y)P(\hat{x}, y)  \rho_x m(\dhx) \mu(\dx)  \right]  \pi(\dy) \\
                &= \int_{\mathscr{Y}} \Tr \!\left[ \left(\int_{\mathscr{X}}   P(x, y)  \rho_x  \mu(\dx) \right) \left( \int_{\mathscr{X}} P(\hat{x}, y) m(\dhx)  \right) \right]  \pi(\dy).
                \label{eq:povm_expected_gain_calculation_3}
        \end{align}
        
    We now further simplify the first factor of the integrand in~\eqref{eq:povm_expected_gain_calculation_3}.
    We have that $\rho_{E}\leq\rho\leq\rho^{\frac{1}{2}}$ for every Borel set $E \in \mathscr{B}(\mathscr{X})$.
    By~\cite[p.~249, Ex.~10.8.8]{schmudgen}, there exists a unique operator $\Omega_{E}$ satisfying
    \begin{equation} \label{eq:Omega_E}
        \left\Vert \Omega_{E}\right\Vert \leq1, \qquad \ker(\rho^{\frac{1}{2}})=\operatorname{ker}(\rho) \subseteq \operatorname{ker}(\Omega_{E}),\qquad \Omega_{E}\rho^{\frac{1}{4}}=\rho_{E}^{\frac{1}{2}}.
    \end{equation}
    Define for $E \in \mathscr{B}(\mathscr{X})$,
    \begin{align}
        \ell(E)\coloneqq \Omega_{E}^{\dag}\Omega_{E}^{\vphantom{\dag}}.
    \end{align}
    Also, recall from the definition of the GPGM that there is a unique contraction $\Lambda_{E}$ satisfying~\eqref{eq:Lambda_E}.
    By the uniqueness of the contraction operator $\Omega_{E}$, we thus have that $\Omega_E= \Lambda_E \rho^{\frac{1}{4}}$, whence $\ell(E)=\rho^{\frac{1}{4}} m^{\PG}(E) \rho^{\frac{1}{4}}$. 
    Therefore, $\ell$ is countably additive on $\mathscr{B}(\mathscr{X})$ under the strong operator topology on $\mathscr{L}_s(\mathscr{H})_+$.
    Moreover, since $-\rho^{\frac{1}{2}} \leq \ell(E) \leq \rho^{\frac{1}{2}}$, it follows by Lemma~\ref{lem:HS_operator_bound_norm} that $\ell(E)$ is a Hilbert--Schmidt operator for all $E \in \mathscr{B}(\mathscr{X})$. Also, we observe that
    \begin{equation}
    \rho^{\frac{1}{4}}\ell(E)\rho^{\frac{1}{4}}=\rho_{E}
    \end{equation}
    for all $E \in \mathscr{B}(\mathscr{X})$.
    By this, we conclude that the first factor in the integrand of~\eqref{eq:povm_expected_gain_calculation_3} is given by
    \begin{equation}
        \int_{\mathscr{X}} P(x, y) \rho_x \mu(\dx) = \int_{\mathscr{X}} P(x, y) \rho^{\frac{1}{4}}\ell(dx)\rho^{\frac{1}{4}}.\label{eq:L_differential}
    \end{equation}
    Using these developments, the expected gain of the POVM, given by~\eqref{eq:povm_expected_gain_calculation_3}, can be further simplified as follows:
    \begin{align}
            G_{\mathscr{E}, S, m} 
                &= \int_{\mathscr{Y}} \Tr\!\left[ \left(\int_{\mathscr{X}} P(x, y) \rho^{\frac{1}{4}}\ell(dx)\rho^{\frac{1}{4}} \right) \left(\int_{\mathscr{X}}  P(\hat{x}, y)    m(\dhx) \right) \right] \pi(\dy)\label{eq:povm_expected_gain_calculation_4} \\
                &= \int_{\mathscr{Y}} \Tr\!\left[ \rho^{\frac{1}{4}}\left(\int_{\mathscr{X}} P(x, y) \ell(dx) \right) \rho^{\frac{1}{4}} \left(\int_{\mathscr{X}}  P(\hat{x}, y)    m(\dhx) \right) \right] \pi(\dy)\label{eq:povm_expected_gain_calculation_13} \\
                &= \int_{\mathscr{Y}} \Tr\!\left[ \left(\int_{\mathscr{X}} P(x, y) \ell(dx) \right) \rho^{\frac{1}{4}}\left(\int_{\mathscr{X}}  P(\hat{x}, y)    m(\dhx)  \right) \rho^{\frac{1}{4}} \right] \pi(\dy)\label{eq:povm_expected_gain_calculation_12} \\
                &= \int_{\mathscr{Y}} \Tr\!\left[ \left(\int_{\mathscr{X}} P(x, y) \ell(dx) \right) \left(\int_{\mathscr{X}}  P(\hat{x}, y)  \rho^{\frac{1}{4}}  m(\dhx) \rho^{\frac{1}{4}} \right) \right] \pi(\dy)\label{eq:povm_expected_gain_calculation_6} \\
                &\leq \int_{\mathscr{Y}} \left\|\int_{\mathscr{X}} P(x, y) \ell(dx) \right\|_2   \left\|\int_{\mathscr{X}}  P(\hat{x}, y)  \rho^{\frac{1}{4}}  m(\dhx) \rho^{\frac{1}{4}} \right\|_2 \pi(\dy)\label{eq:povm_expected_gain_calculation_7} \\
                &\leq \sqrt{\left(\int_{\mathscr{Y}} \left\|\int_{\mathscr{X}} P(x, y) \ell(dx) \right\|_2^2 \pi(\dy) \right) \left(\int_{\mathscr{Y}} \left\|\int_{\mathscr{X}}  P(\hat{x}, y)  \rho^{\frac{1}{4}}  m(\dhx) \rho^{\frac{1}{4}} \right\|_2^2 \pi(\dy) \right) }   \label{eq:povm_expected_gain_calculation_8} \\
                &= \sqrt{\int_{\mathscr{Y}} \Tr\!\left[ \left(\int_{\mathscr{X}} P(x, y) \ell(dx) \right) \left(\int_{\mathscr{X}} P(x', y) \ell(dx') \right) \right] \pi(\dy)} \nonumber \\
                &\hspace{0.5cm} \times \sqrt{ \int_{\mathscr{Y}}\Tr\!\left[ \left(\int_{\mathscr{X}}  P(\hat{x}, y)  \rho^{\frac{1}{4}}  m(\dhx) \rho^{\frac{1}{4}} \right) \left(\int_{\mathscr{X}}  P(\hat{x}', y)  \rho^{\frac{1}{4}}  m(\dhx') \rho^{\frac{1}{4}} \right) \right] \pi(\dy)}. \label{eq:povm_expected_gain_calculation_9}
        \end{align}
    The equality~\eqref{eq:povm_expected_gain_calculation_4} follows from~\eqref{eq:L_differential},  and~\eqref{eq:povm_expected_gain_calculation_12} uses the cyclicity of trace
    {In~\eqref{eq:povm_expected_gain_calculation_13} we are allowed to take the operators $\rho^{\frac{1}{4}}$ outside the first integral, and}
    {in~\eqref{eq:povm_expected_gain_calculation_6} we are allowed to take the operators $\rho^{\frac14}$ inside the second inner integral, as one can see by writing down the definition of Bochner integral directly in terms of simple measurable functions.} 
    Also,~\eqref{eq:povm_expected_gain_calculation_7} and~\eqref{eq:povm_expected_gain_calculation_8} are Cauchy--Schwarz inequalities, the former at the operator level and the latter at the level of functions.
    By Theorem~\ref{thm:int_hilbert_schmidt_valued_measure}, we can take the trace function inside the integrals in~\eqref{eq:povm_expected_gain_calculation_9} to conclude that
    \begin{align}
            G_{\mathscr{E}, S, m} 
                &\leq \sqrt{ \int_{\mathscr{Y}}  \int_{\mathscr{X}\times \mathscr{X}} P(x, y) P(x', y) \Tr\!\left[\ell(dx)    \ell(dx')\right]  \pi(\dy) } \nonumber \\
                &\hspace{0.5cm} \times \sqrt{ \int_{\mathscr{Y}}  \int_{\mathscr{X} \times \mathscr{X}} P(\hat{x}, y) P(\hat{x}', y)  \Tr\!\left[\rho^{\frac{1}{4}}  m(\dhx)     \rho^{\frac{1}{2}}  m(\dhx') \rho^{\frac{1}{4}}  \right]\pi(\dy) } .\label{eq:povm_expected_gain_calculation_10}
        \end{align}
    By Fubini's theorem~\cite[Theorem~5.2.2]{cohn2013measure}, the order of the integrations in~\eqref{eq:povm_expected_gain_calculation_10} can be interchanged to get
    \begin{align}
            G_{\mathscr{E}, S, m} 
                &\leq \sqrt{ \int_{\mathscr{X}\times \mathscr{X}} \left( \int_{\mathscr{Y}} P(x, y) P(x', y)  \pi(\dy)\right)\Tr\!\left[\ell(dx)    \ell(dx')\right]  } \nonumber \\
                &\hspace{0.5cm} \times \sqrt{ \int_{\mathscr{X}\times \mathscr{X}} \left( \int_{\mathscr{Y}} P(\hat{x}, y) P(\hat{x}', y) \pi(\dy) \right) \Tr\!\left[\rho^{\frac{1}{4}}  m(\dhx)     \rho^{\frac{1}{2}}  m(\dhx') \rho^{\frac{1}{4}}  \right] } \\
                &=\sqrt{ \int_{\mathscr{X}\times \mathscr{X}} S(x, x')\Tr\!\left[\ell(dx)    \ell(dx')\right]  } \nonumber \\
                &\hspace{0.5cm} \times \sqrt{ \int_{\mathscr{X}\times \mathscr{X}} S(\hat{x}, \hat{x}') \Tr\!\left[\rho^{\frac{1}{4}}  m(\dhx)     \rho^{\frac{1}{2}}  m(\dhx') \rho^{\frac{1}{4}}  \right] } .\label{eq:povm_expected_gain_calculation_11}
        \end{align}
    Here we used the fact that $S=P \star_{\pi} P$.
    Now, recall that $\ell(E)=\rho^{\frac{1}{4}} m^{\PG}(E) \rho^{\frac{1}{4}}$ and $\rho^{\frac{1}{2}} m^{\PG}(E) \rho^{\frac{1}{2}}=\rho_E$ for all $E \in \mathscr{B}(\mathscr{X})$.
    This gives 
    \begin{align}
        &\int_{\mathscr{X}\times \mathscr{X}} S(x, x')  \Tr\!\left[\ell(dx)  \ell(dx')\right]   \notag \\
        &= \int_{\mathscr{X}\times \mathscr{X}} S(x, x') \Tr\!\left[\rho^{\frac{1}{4}}m^{\PG}(dx)\rho^{\frac{1}{4}}\rho^{\frac{1}{4}}m^{\PG}(dx^{\prime})\rho^{\frac{1}{4}}\right] \label{eq:differential_Trace_double_L_2}\\
        &=\int_{\mathscr{X}\times \mathscr{X}} S(x, x')  \Tr\!\left[m^{\PG}(dx)\rho^{\frac{1}{2}}m^{\PG}(dx^{\prime})\rho^{\frac{1}{2}}\right]  \\
        &=\int_{\mathscr{X}} \int_{ \mathscr{X}} S(x, x')  \Tr\!\left[m^{\PG}(dx)\rho_{x^{\prime}}\right]\mu(dx^{\prime}).\label{eq:differential_Trace_double_L}
    \end{align}
    By substituting the equality~\eqref{eq:differential_Trace_double_L} into~\eqref{eq:povm_expected_gain_calculation_11}, we conclude that
    \begin{align}
            G_{\mathscr{E}, S, m} 
                &\leq \sqrt{\int_{\mathscr{X}} \int_{ \mathscr{X}} S(x, x')\Tr\!\left[m^{\PG}(dx)\rho_{x^{\prime}}\right]\mu(dx^{\prime})  } \nonumber \\
                &\hspace{0.5cm} \times \sqrt{ \int_{\mathscr{X}\times \mathscr{X}} S(\hat{x}, \hat{x}') \Tr\!\left[\rho^{\frac{1}{4}}  m(\dhx)     \rho^{\frac{1}{2}}  m(\dhx') \rho^{\frac{1}{4}}  \right] } \\
                &\leq \sqrt{ \int_{\mathscr{X}} \int_{ \mathscr{X}} S(x, x')\Tr\!\left[\rho_{x^{\prime}} m^{\PG}(dx)\right]\mu(dx^{\prime})  } \nonumber \\
                &\hspace{0.5cm} \times \sqrt{ \int_{\mathscr{X}\times \mathscr{X}} \Tr\!\left[\rho^{\frac{1}{4}}  m(\dhx)     \rho^{\frac{1}{2}}  m(\dhx') \rho^{\frac{1}{4}}  \right] } \\
                &= \sqrt{ \int_{\mathscr{X}} \Tr\!\left[\rho_{x^{\prime}} \int_{\mathscr{X}} S(x, x') m^{\PG}(dx)\right]\mu(dx^{\prime})  }\nonumber \\
                &\hspace{0.5cm} \times \sqrt{  \Tr\!\left[\rho^{\frac{1}{4}}  m(\mathscr{X})     \rho^{\frac{1}{2}}  m(\mathscr{X}) \rho^{\frac{1}{4}}  \right] } \\
                &= \sqrt{ \int_{\mathscr{X}} \Tr\!\left[\rho_{x^{\prime}} \int_{\mathscr{X}} S(x, x') m^{\PG}(dx)\right]\mu(dx^{\prime})  }\  \sqrt{  \Tr\!\left[\rho^{\frac{1}{4}}     \rho^{\frac{1}{2}} \rho^{\frac{1}{4}}  \right] } \\
                &= \sqrt{ \int_{\mathscr{X}} \Tr\!\left[\rho_{x^{\prime}} \int_{\mathscr{X}} S(x, x') m^{\PG}(dx)\right]\mu(dx^{\prime})  }\\
                &= \sqrt{G_{\mathscr{E}, S, m^{\PG}}}.
        \end{align}
    This concludes the proof. 
    \end{proof}

\medskip 

    As a direct consequence of the main result, we conclude the following well-known result.
    \Frame{
    \begin{corollary}[Barnum--Knill theorem]
        The optimal success probability of discrimination of the states of a discrete quantum ensemble does not exceed the square root of the corresponding success probability of the pretty good measurement; i.e., the inequality~\eqref{eq:bk} holds.
    \end{corollary}
    }
    \begin{proof}
        It follows directly from Theorem~\ref{thm:generalized_bk} by taking $\mathscr{X}$ to be a finite set, $\mu$ to be 
        any probability distribution on $\mathscr{X}$, and the positive score function $S$ to be the identity matrix, i.e., $S(x,y)=\delta_{xy}$ for $x,y \in \mathscr{X}$.
    \end{proof}

\subsection{Bayesian estimation of parameters for ensembles}

    \label{sec:mse}

    The main aim of the section is to show that the \emph{mean square error} (MSE) of the GPGM is nearly optimal for Bayesian estimation.
    
\subsubsection{Mean square error of a measurement}

    Consider 
    a quantum ensemble on a complex separable Hilbert space $\mathscr{H}$ given by $\mathscr{E} \coloneqq \left(\left(\mu(\dx), \rho_x \right)\right)_{x\in \mathds{R}^N}$. 
    Here, $\rho_x\in \mathscr{D}(\mathscr{H})$ for each $x \in \mathds{R}^{N}$, and $\mu$ is a Borel probability measure on $\mathds{R}^{N}$. 
    The primary case of interest for applications is as considered in \cite{holevo_accessible_2021,mlmw2024}, when the underlying quantum system is a bosonic system with $n$ modes, so that $\mathscr{H} \simeq L^2(\mathds{R}^n)$, $N=2n$, and $\rho_x = D(x) \rho_0 D(-x)$, with $\rho_0$ a fixed state and $D(x)$ the Weyl displacement operator corresponding to the vector $x\in\mathds{R}^{2n}$. However, our results are much more general, and apply to any ensemble parameterized by elements of a Euclidean space.

    A quantum measurement $m \in \mathscr{M}(\mathds{R}^{N}, \mathscr{L}_s(\mathscr{H})_+)$ yields an estimate $\hat{x}$ of the true parameter~$x$ of the quantum system. 
    For a given true parameter $x$ of the system, we know that each estimate $\hat{x}$, resulting from the measurement $m$, is distributed according to the probability density $\Tr\!\left[ \rho_x m(\dhx) \right]$, given by~\eqref{eq:cond_probability_POVM}.
    The MSE of the measurement $m$ is the following function of the parameter $x$:
    \begin{align}
        \operatorname{E}_{m}(x) \coloneqq \int_{\mathds{R}^{N}} \left\|x-\hat{x}\right\|^2 \Tr\!\left[ \rho_x m(\dhx) \right].
    \end{align}
    The expected MSE of the measurement is given by
    \begin{align}
        \MSE(\mathscr{E},m)
            \coloneqq&\ \int_{\mathds{R}^{N}} \operatorname{E}_{m}(x) \mu(\dx) \\
            =&\ \int_{\mathds{R}^{N}} \int_{\mathds{R}^{N}} \left\|x-\hat{x}\right\|^2 \Tr\!\left[ \rho_x m(\dhx) \right]  \mu(\dx). \label{eq:exp_mse}
    \end{align}
    
    The following lemma states that the expected MSE of a measurement can be expressed as a limit of the expected gain of the measurement, with respect to a specific type of Gaussian score function.
    \Frame{
    \begin{lemma}
        \label{lem:mse_limit_gain_relation}
        For 
        an ensemble $\mathscr{E} \coloneqq \left(\left(\mu(\dx), \rho_x \right)\right)_{x\in \mathds{R}^N}$ over a complex separable Hilbert space~$\mathscr{H}$ and a quantum measurement $m$, the following equality holds:
        \begin{align}
            \label{eq:mse_limit_exp_gain_relation}
            \MSE(\mathscr{E},m) &= \lim_{t \searrow 0} \frac{2}{t}\left(1-G_{\mathscr{E}, S_{tI}, m}\right),
        \end{align}
        where {the Gaussian score function $S_{\Sigma}$ is constructed in Definition~\ref{def:gaussian_score_function}, and here $\Sigma=tI$, with $t>0$ and $I$ denoting} the $N \times N$ identity matrix.
    \end{lemma}
    }
    \begin{proof}
        Recall from~\eqref{eq:exp_gain} and~\eqref{eq:exp_gain_trace_inside_integral} that the expected gain of a measurement $m$ for the ensemble $\mathscr{E}$, corresponding to the Gaussian score function $S_{tI}$, is given by
        \begin{align}
            G_{\mathscr{E}, S_{tI}, m}
                &= \int_{\mathds{R}^{N}} \int_{\mathds{R}^{N}} \exp\!\left(-\frac{t}{2} \left\|x-\hat{x}\right\|^2\right) \Tr \!\left[\rho_x m(\dhx) \right] \mu(\dx) .
        \end{align}
        This gives
        \begin{align}
            \frac{2}{t}\left(1-G_{\mathscr{E}, S_{tI}, m}\right)
                &= \int_{\mathds{R}^{N}} \int_{\mathds{R}^{N}} \frac{2}{t} \left(1- \exp\!\left(-\frac{t}{2} \left\|x-\hat{x}\right\|^2\right) \right) \Tr \!\left[\rho_x m(\dhx) \right] \mu(\dx).
        \end{align}
        For $t > 0$, by integrating the elementary inequality
        \begin{equation}
            \frac{2}{t} \left(1- \exp\!\left(-\frac{t}{2} \left\|x-\hat{x}\right\|^2\right) \right) \leq \left\|x-\hat{x}\right\|^2, \qquad \forall x, \hat{x} \in \mathds{R}^{N},
        \end{equation}
        we obtain the relation
        \begin{align}
            \MSE(\mathscr{E},m)  \geq \frac{2}{t}\left(1-G_{\mathscr{E}, S_{tI}, m}\right).
        \end{align}
        Therefore, we get
        \begin{multline}
            \label{eq:mse_limsup}
            \MSE(\mathscr{E},m) \\
            \geq \limsup_{t \searrow 0} \int_{\mathds{R}^{N}} \int_{\mathds{R}^{N}} \frac{2}{t} \left(1- \exp\!\left(-\frac{t}{2} \left\|x-\hat{x}\right\|^2\right) \right) \Tr \!\left[\rho_x m(\dhx) \right] \mu(\dx).
        \end{multline}
        Also, we know that
        \begin{align}
            \left\|x-\hat{x}\right\|^2 = \lim_{t \searrow 0} \frac{2}{t} \left(1- \exp\!\left(-\frac{t}{2} \left\|x-\hat{x}\right\|^2\right) \right).
        \end{align}
        By Fatou's lemma~\cite[Theorem~2.4.4]{cohn2013measure}, we thus get
        \begin{align}
            \label{eq:mse_liminf}
            \MSE(\mathscr{E},m) \leq \liminf_{t \searrow 0} \int_{\mathds{R}^{N}} \int_{\mathds{R}^{N}} \frac{2}{t} \left(1- \exp\!\left(-\frac{t}{2} \left\|x-\hat{x}\right\|^2\right) \right) \Tr \!\left[\rho_x m(\dhx) \right] \mu(\dx).
        \end{align}
        By~\eqref{eq:mse_limsup} and~\eqref{eq:mse_liminf}, we conclude that the limit in~\eqref{eq:mse_limit_exp_gain_relation} holds.
    \end{proof}

    The expected MSE of the GPGM is finite under the condition that $\mu$ has a  \emph{finite second moment}, as stated in the following proposition.
    Recall that the second moment of a probability measure $\mu$ on $\mathds{R}^N$ is defined by
    \begin{align}
        \operatorname{E}_{\mu, 2} \coloneqq \int_{\mathds{R}^{N}} \left\|x\right\|^2 \mu(\dx).
    \end{align}
    \Frame{
    \begin{proposition}
        Let $\mathscr{E} \coloneqq \left(\left(\mu(\dx), \rho_x \right)\right)_{x\in \mathds{R}^N}$ be a quantum ensemble over a complex separable Hilbert space.
        The expected mean square error of the generalized pretty good measurement of the ensemble, as given by Definition~\ref{def:pgm}, is finite if the second moment of $\mu$ is finite.
    \end{proposition}
    }
    \begin{proof}
        Consider the expected MSE of the GPGM given by \eqref{eq:exp_mse}:
        \begin{align}
            \MSE(\mathscr{E}, m^{\PG} ) 
                &= \int_{\mathds{R}^N} \int_{\mathds{R}^N} \left\|x-\hat{x}\right\|^2 \Tr\!\left[ \rho_x m^{\PG}(\dhx) \right]  \mu(\dx) \\
                &= \int_{\mathds{R}^N} \int_{\mathds{R}^N} \left(\left\|x\right\|^2+\left\|\hat{x}\right\|^2-2 x^\intercal\hat{x} \right) \Tr\!\left[ \rho_x m^{\PG}(\dhx) \right]  \mu(\dx) \\
                &\leq \int_{\mathds{R}^N} \int_{\mathds{R}^N} \left(\left\|x\right\|^2+\left\|\hat{x}\right\|^2+2 \left\|x\right\|\left\|\hat{x}\right\| \right) \Tr\!\left[ \rho_x m^{\PG}(\dhx) \right]  \mu(\dx) \\
                &\leq  \int_{\mathds{R}^N} \int_{\mathds{R}^N} 2\left(\left\|x\right\|^2+\left\|\hat{x}\right\|^2\right) \Tr\!\left[ \rho_x m^{\PG}(\dhx) \right]  \mu(\dx) \\
                &=  \int_{\mathds{R}^N} \int_{\mathds{R}^N} 2\left\|x\right\|^2 \Tr\!\left[ \rho_x m^{\PG}(\dhx) \right]  \mu(\dx) \nonumber \\
                &\hspace{2cm}+  \int_{\mathds{R}^N} \int_{\mathds{R}^N} 2\left\|\hat{x}\right\|^2 \Tr\!\left[ \rho_x m^{\PG}(\dhx) \right]  \mu(\dx)  \\
                &= \int_{\mathds{R}^N} 2\left\|x\right\|^2 \left( \int_{\mathds{R}^N}  \Tr\!\left[ \rho_x m^{\PG}(\dhx) \right] \right) \mu(\dx) \nonumber \\
                &\hspace{2cm}+  \int_{\mathds{R}^N} \int_{\mathds{R}^N}2 \left\|\hat{x}\right\|^2 \Tr\!\left[ \rho_x m^{\PG}(\dhx) \right]  \mu(\dx) \\
                &= 2  \operatorname{E}_{\mu, 2}+  \int_{\mathds{R}^N} \int_{\mathds{R}^N} 2\left\|\hat{x}\right\|^2 \Tr\!\left[ \rho_x m^{\PG}(\dhx) \right]  \mu(\dx). \label{eq:mse_finite_calculation}
        \end{align}
        {The above derivation holds for a generic POVM, as we have not used the specific properties of $m^{\PG}$ so far. We now make use of the fact that $m^{\PG}$ is the generalized pretty good measurement to note that} the second term in~\eqref{eq:mse_finite_calculation} is equal to the first term.
        Indeed, consider the operator-valued measure $\ell$ defined in the proof of Theorem~\ref{thm:generalized_bk}.
        By similar arguments as presented in the development~\eqref{eq:differential_Trace_double_L_2}--\eqref{eq:differential_Trace_double_L}, we conclude that
        \begin{align}
            & \int_{\mathds{R}^N} \int_{\mathds{R}^N} \left\|\hat{x}\right\|^2 \Tr\!\left[ \rho_x m^{\PG}(\dhx) \right]  \mu(\dx)\notag \\
                &= \int_{\mathds{R}^N} \int_{\mathds{R}^N} \left\|\hat{x}\right\|^2 \Tr\!\left[ \ell(\dx) \ell(\dhx) \right] \\
                &= \int_{\mathds{R}^N} \int_{\mathds{R}^N} \left\|\hat{x}\right\|^2 \Tr\!\left[ \ell(\dhx) \ell(\dx) \right] \\
                &= \int_{\mathds{R}^N} \int_{\mathds{R}^N} \left\|\hat{x}\right\|^2 \Tr\!\left[ \rho_{\hat{x}} m^{\PG}(\dx) \right]  \mu(\dhx) \\
                &= \int_{\mathds{R}^N} \left\|\hat{x}\right\|^2 \left( \int_{\mathds{R}^N}  \Tr\!\left[ \rho_{\hat{x}} m^{\PG}(\dx) \right] \right) \mu(\dhx) \\
                &=   \operatorname{E}_{\mu, 2}.
        \end{align}
        We thus obtain from~\eqref{eq:mse_finite_calculation} that
        \begin{align}
            \MSE(\mathscr{E}, m^{\PG} )  \leq 4 \operatorname{E}_{\mu, 2} < \infty,
        \end{align}
        concluding the proof.
    \end{proof}

\subsubsection{Near-optimal mean square error of generalized pretty good measurement}

    The following result states that the expected MSE of the GPGM is not more than twice the minimal expected MSE.
    \Frame{
    \begin{theorem}
        \label{thm:mse}
        For an arbitrary quantum ensemble $\mathscr{E} \coloneqq \left(\left(\mu(\dx), \rho_x \right)\right)_{x\in \mathds{R}^N}$ and quantum measurement $m$, the following inequality holds:
        \begin{align}
            \label{eq:mse_pgm_inequality}
            \MSE(\mathscr{E}, m^{\PG}) \leq 2\MSE(\mathscr{E},m).
        \end{align}
    \end{theorem}
    } 
    \begin{proof}
        Let $t>0$ be arbitrary, and let $I$ denote the $N \times N$ identity matrix.
        We know that $0 \leq G_{\mathscr{E}, S_{tI}, m} \leq 1${, where the Gaussian score function is given in Definition~\ref{def:gaussian_score_function}}.
        By Theorem~\ref{thm:generalized_bk}, we thus conclude that
        \begin{align}
            \frac{2}{t}  \left(1- G_{\mathscr{E}, S_{tI}, m^{\PG}}\right)
                &\leq \frac{2}{t}  \left(1- G^2_{\mathscr{E}, S_{tI}, m}\right) \label{eq:gain_square_pgm_inequality}\\
                &= \frac{2}{t} \left(1+ G_{\mathscr{E}, S_{tI}, m}\right) \left(1- G_{\mathscr{E}, S_{tI}, m}\right) \\
                &\leq 2 \left[\frac{2}{t} \left(1- G_{\mathscr{E}, S_{tI}, m}\right)\right].
                    \label{eq:gain_square_pgm_inequality_2}
        \end{align}
        By taking the limit $t \searrow 0$ in~\eqref{eq:gain_square_pgm_inequality_2} and applying Lemma~\ref{lem:mse_limit_gain_relation}, we thus arrive at the desired inequality~\eqref{eq:mse_pgm_inequality}.
    \end{proof}

     Theorem~\ref{thm:mse} establishes near-optimality of the GPGM in the Bayesian estimation of unknown parameters of a quantum ensemble. 
     As a special case of this result, the GPGM for bosonic Gaussian ensembles, previously studied in~\cite{holevo2020gaussian, holevo2021classical, holevo_accessible_2021, mlmw2024}, indeed does pretty well for the parameter estimation task.
     Since the GPGM of a bosonic Gaussian ensemble is known to be Gaussian~\cite{holevo2020gaussian} and its MSE is explicitly available~\cite{mlmw2024}, it is a desirable theoretical and  practical tool in Bayesian estimation of unknown parameters in bosonic Gaussian ensembles.

\section{Conclusion}

    We have extended the notion of pretty good measurement to general alphabets and arbitrary (possibly infinite-dimensional) quantum ensembles.
    We also introduced a notion called the expected gain of a measurement on an ensemble, an instance of which is the success probability of the measurement in the quantum state discrimination task.
    Our main result is a generalization of the Barnum--Knill theorem, 
    in which we have established that the square root of the expected gain of the
    generalized pretty good measurement  is 
    no smaller than the optimal expected gain.
    We also show that the expected 
    mean square error of the generalized pretty good measurement is 
    no more than twice the least expected 
    mean square error on the ensemble.

\section*{Acknowledgment}

    HKM acknowledges support from the NSF under grant number 2304816 and AFRL under agreement number FA8750-23-2-00.
    HKM thanks Komal Malik for insightful discussions.
    LL~acknowledges financial support from MIUR (Ministero dell'Istruzione, dell'Universit\`{a} e della Ricerca) through the project `Dipartimenti di Eccellenza 2023--2027' of the `Classe di Scienze' department at the Scuola Normale Superiore, and from the European Union under the European Research Council (ERC grant agreement number 101165230).
    MMW acknowledges support from the NSF under grant number 2329662. 

\begin{Backmatter}

\bibliographystyle{unsrtnat}
\bibliography{reference}
\end{Backmatter}
\printaddress

\appendices

\section{Integration with respect to operator-valued measures}
    \label{app:int_wrt_povm}
    Let $\mathscr{X}$ be a locally compact Hausdorff space, let $\mu$ be a Borel probability measure on $\mathscr{X}$,
    and let $\mathscr{H}$ be a complex separable Hilbert space.
    Suppose $\ell$ is a positive operator-valued measure, i.e., an operator-valued measure on $\mathscr{X}$ such that $\ell(E) \in \mathscr{L}_s(\mathscr{H})_+$ for all $E \in \mathscr{B}(\mathscr{X})$.
    In what follows, we will discuss the integration of real-valued functions with respect to positive operator-valued measures.\footnote{In~\cite{Mitter_etal_1984}, the authors work with an extra assumption on self-adjoint operator-valued measures to be \emph{regular}. 
    This assumption is mainly needed when working with bounded linear maps $L\colon C_0(\mathscr{X}) \to \mathscr{L}(X, Z^*)$, where $C_0(\mathscr{X})$ is the space of real-valued continuous maps on $\mathscr{X}$ that vanish at infinity, and $X, Z$ are Banach spaces. 
    The theory of integration can be developed without the regularity assumption.
    {Also, the assumption that $\mathscr{X}$ be a locally compact Hausdorff space is merely choice for its relevance to the quantum estimation theory; the integration theory can be developed with a general measurable space.}
    See the first paragraph of Section~4 in~\cite{Mitter_etal_1984}.}
    We refer the reader to~\cite{Mitter_etal_1984} for a detailed theory of integration involving operator-valued measures.

    Let $f\colon \mathscr{X} \to \mathds{R}$ be a real-valued simple measurable function given by
    \begin{align}
        f \coloneqq \sum_{i=1}^n \mathbf{1}_{E_i} \otimes \alpha_i,
    \end{align}
    where $\mathbf{1}_E$ denotes the indicator function for any subset $E\subseteq \mathscr{X}$, $E_1,\ldots, E_n$ are disjoint Borel subsets of $\mathscr{X}$, and $\alpha_1,\ldots, \alpha_n \in \mathds{R}$.
    The integration of $f$ with respect to $\ell$ is defined unambiguously by
    \begin{align}
        \int_{\mathscr{X}} f(x) \ell(\dx) \coloneqq \sum_{i=1}^n \ell(E_i) \alpha_i.
    \end{align}
    Let $L_\infty(\mathscr{X})$ denote the Banach space of bounded, real-valued Borel measurable functions on $\mathscr{X}$ equipped with the uniform norm.
    It is well known that every $f \in L_\infty(\mathscr{X})$ is the uniform limit of a sequence of real-valued simple measurable functions on $\mathscr{X}$.
    Define the \emph{operator norm semivariation} of $\ell$ as
    \begin{align}
        \label{eq:norm_semivariation}
       \left\| \ell \right\| \coloneqq \sup \bigg\{\left\| \int_{\mathscr{X}} f(x) \ell(\dx) \right\|: \|f\|_\infty \leq 1, f\colon \mathscr{X}\to \mathds{R} \text{ simple} \bigg\}=\|\ell(\mathscr{X})\|.
    \end{align}
    To show the last equality, note first that, for any simple measurable function $f\colon \mathscr{X} \to \mathds{R}$, we have
    \begin{align}
        -\|f\|_\infty \ell(\mathscr{X}) \leq \int_{\mathscr{X}} f(x) \ell(\dx) \leq \|f\|_\infty \ell(\mathscr{X}).
    \end{align}
    It thus follows that $\left\| \ell \right\| \leq 
    \|\ell(\mathscr{X})\|$. The opposite inequality is obtained by considering $f$ identically equal to $1$.
    
    This leads to the following definition of integration of $f \in L_\infty(\mathscr{X})$ with respect to $\ell$ as
    \begin{align}\label{eq:int_wrt_povm_real_valued_maps}
        \int_{\mathscr{X}} f(x) \ell(\dx) \coloneqq \lim_{n \to \infty} \int_{\mathscr{X}} f_n(x) \ell(\dx),
    \end{align}
    where $(f_n)_{n \in \mathds{N}}$ is an arbitrary sequence of real-valued simple measurable functions on $\mathscr{X}$ converging to $f$ uniformly, and the limit in~\eqref{eq:int_wrt_povm_real_valued_maps} is with respect to the operator norm topology. We emphasize that said limit does not depend on the choice of the sequence $(f_n)_{n \in \mathds{N}}$, and hence the integration is well defined.
    Indeed, if $(g_n)_{n \in \mathds{N}}$ is any other sequence of real-valued simple measurable functions on $\mathscr{X}$ that uniformly converges to $f$ then we have
    \begin{align}
        \left\|\int_{\mathscr{X}} f_n(x) \ell(\dx) - \int_{\mathscr{X}} g_n(x) \ell(\dx) \right\|\label{eq:int_wrt_povm_real_valued_simple_function_1}
            &= \left\|\int_{\mathscr{X}} \left(f_n(x) -  g_n(x) \right) \ell(\dx) \right\| \\
            &\leq \left\| \ell \right\| \|f_n - g_n \|_\infty.
    \end{align}
    Since $\left\| \ell \right\|=\|\ell(\mathscr{X})\| < \infty$, we thus get
    \begin{align}
        \lim_{n \to \infty} \int_{\mathscr{X}} f_n(x) \ell(\dx) = \lim_{n \to \infty} \int_{\mathscr{X}} g_n(x) \ell(\dx). \label{eq:int_wrt_povm_real_valued_simple_function_2}
    \end{align}
    See Section~3 of~\cite{Mitter_etal_1984} for the development of the integration theory of real-valued maps with respect to operator-valued measures in a more general setting.
    
    Observe that the integration of $f \in L_\infty(\mathscr{X})$ with respect to $\ell$ is an element of $\mathscr{L}_s(\mathscr{X})$.
    To recapitulate, this is mainly due to the following reasons:
    \begin{enumerate}[label=(\roman*)]
        \item  $\ell$ takes values $\mathscr{L}_s(\mathscr{X})_+$,
        \item  the integration of simple measurable functions with respect to $\ell$ is an element of $\mathscr{L}_s(\mathscr{X})$,
        \item $\ell$ has finite operator norm semivariation given by~\eqref{eq:norm_semivariation}, which implies that the limit in~\eqref{eq:int_wrt_povm_real_valued_maps} exists in the operator norm.
    \end{enumerate}
    Furthermore, it is clear by the definition that, if $f$ is a non-negative function, then its integration with respect to $\ell$ is an element of $\mathscr{L}_s(\mathscr{H})_+$.
    The above observation leads to the following two results, which are interesting on their own. 
    \Frame{
    \begin{theorem}\label{thm:int_trace_class_valued_measure}
        Suppose $\ell(E) \in \mathscr{L}_{st}(\mathscr{H})_+$ for all $E \in \mathscr{B}(\mathscr{X})$.
        Then, for $f \in L_\infty(\mathscr{X})$, we have
        \begin{align}
            \int_\mathscr{X} f(x) \ell(\dx) \in \mathscr{L}_{st}(\mathscr{H}).
        \end{align}
        Furthermore, if $f$ is a non-negative function, then 
        \begin{align}
            \int_\mathscr{X} f(x) \ell(\dx) \in \mathscr{L}_{st}(\mathscr{H})_+.
        \end{align}
        Also, we have
        \begin{align}\label{eq:int_trace_trace_int}
            \Tr\!\left[\int_\mathscr{X} f(x) \ell(\dx) \right] = \int_\mathscr{X} f(x) \Tr\!\left[\ell(\dx)\right].
        \end{align}
        Here $\Tr\!\left[\ell(\cdot)\right]$ is the finite positive measure on $\mathscr{X}$ given by $\mathscr{B}(\mathscr{X}) \ni E \mapsto \Tr\!\left[\ell(E)\right]$.
    \end{theorem}
    }
    \begin{proof}
        Let $\ell$ be an operator-valued measure such that $\ell(E)\in \mathscr{L}_{st}(\mathscr{H})_+$ for all $E \in \mathscr{B}(\mathscr{X})$, and let $f \in L_\infty(\mathscr{X})$.
        Suppose $(f_n)_{n \in \mathds{N}}$ is a sequence of real-valued simple measurable functions on $\mathscr{X}$ converging to $f$ uniformly so that
        \begin{align}\label{eq:int_wrt_povm_real_valued_maps_oper_norm}
            \int_\mathscr{X} f(x) \ell(\dx) = \lim_{n \to \infty} \int_{\mathscr{X}} f_n(x) \ell(\dx).
        \end{align}
        The above general construction, which works for measures taking on values in the space of bounded operators, guarantees that the limit in~\eqref{eq:int_wrt_povm_real_valued_maps_oper_norm} exists with respect to the operator norm topology. However, in our current setting in which the measure $\ell$ takes on values in the space of trace-class operators, it is possible to show that the limit in~\eqref{eq:int_wrt_povm_real_valued_maps_oper_norm} actually exists in the trace norm topology. To this end, start by observing that $\int_{\mathscr{X}} f_n(x) \ell(\dx) \in \mathscr{L}_{st}(\mathscr{H})$ for all $n \in \mathds{N}$.

        It is easily seen for any simple measurable function $g$ that the integral $\int_{\mathscr{X}} g(x) \ell(\dx)$ is a self-adjoint trace-class operator.
        Moreover, we have
        \begin{align}
            -\|g\|_\infty \ell(\mathscr{X}) \leq \int_{\mathscr{X}} g(x) \ell(\dx) \leq \|g\|_\infty \ell(\mathscr{X}),
        \end{align}
        which implies that
        \begin{align}
            \left\| \int_{\mathscr{X}} g(x) \ell(\dx)  \right\|_1 \leq \left\|g\right\|_\infty \left\|\ell(\mathscr{X})\right\|_1.
        \end{align}
        Consequently, the \emph{trace norm semivariation} of $\ell$ is finite:
        \begin{align}
        \label{eq:trace_norm_variation_m}
            \left\| \ell \right\|_1 \coloneqq \sup \bigg\{\left\| \int_{\mathscr{X}} g(x) \ell(\dx) \right\|_1: \|g\|_\infty \leq 1, g\colon \mathscr{X}\to \mathds{R} \text{ simple} \bigg\} = \left\|\ell(\mathscr{X})\right\|_1 < \infty.
        \end{align}
        We have for $n, n' \in \mathds{N}$
        \begin{align}
            \left\| \int_{\mathscr{X}} f_n(x) \ell(\dx)-\int_{\mathscr{X}} f_{n'}(x) \ell(\dx) \right\|_1
                & \leq \left\| \ell \right\|_1 \left\| f_n - f_{n'} \right\|_\infty,
        \end{align}
        thus implying that the sequence $\left(\int_{\mathscr{X}} f_n(x) \ell(\dx) \right)_{n \in \mathds{N}}$ is Cauchy in the trace norm.
        Lastly, the fact that $\left\|\cdot \right\| \leq \left\| \cdot \right\|_1$, combined with~\eqref{eq:int_wrt_povm_real_valued_maps_oper_norm}, implies that the sequence converges to $\int_{\mathscr{X}} f(x) \ell(\dx)$ in the trace norm.
        Consequently, $\int_{\mathscr{X}} f(x) \ell(\dx) \in \mathscr{L}_{st}(\mathscr{H})$.
        Assume that $f$ is a non-negative function. 
        Then there exists a sequence of non-negative simple measurable functions converging to $f$ uniformly.
        Since the integration of a non-negative simple measurable function with respect to $\ell$ is an element of $\mathscr{L}_{st}(\mathscr{H})_+$, we have $\int_{\mathscr{X}} f(x) \ell(\dx) \in \mathscr{L}_{st}(\mathscr{H})_+$.

        Now, we have for all $n \in \mathds{N}$
        \begin{align}
            \Tr\!\left[  \int_{\mathscr{X}} f_n(x) \ell(\dx)\right] =  \int_{\mathscr{X}} f_n(x) \Tr\!\left[\ell(\dx)\right].
        \end{align}
        From the continuity of the trace operator with respect to the trace norm and by taking the limit $n \to \infty$, we thus obtain that
        \begin{align}\label{eq:trace_int}
            \Tr\!\left[  \int_{\mathscr{X}} f(x) \ell(\dx)\right] = \lim_{n \to \infty} \int_{\mathscr{X}} f_n(x) \Tr\!\left[\ell(\dx)\right].
        \end{align}
        We know that $f$, being a bounded measurable function, is integrable with respect to the finite positive measure $\Tr\!\left[\ell(\cdot)\right]$. 
        Moreover,
        \begin{align}\label{eq:int_wrt_trace_measure}
             \int_{\mathscr{X}} f(x) \Tr\!\left[\ell(\dx)\right]= \lim_{n \to \infty} \int_{\mathscr{X}} f_n(x) \Tr\!\left[\ell(\dx)\right].
        \end{align}
        From~\eqref{eq:trace_int} and~\eqref{eq:int_wrt_trace_measure}, we thus get the desired equality~\eqref{eq:int_trace_trace_int}.
    \end{proof}

    The following lemma will be useful in the proof of the next result.
    \Frame{
    \begin{lemma}
        \label{lem:HS_operator_bound_norm}
        Let $\varrho \in \mathscr{L}_{s}(\mathscr{H})$ and $\zeta \in \mathscr{L}_{s\operatorname{HS}}(\mathscr{H})_+$ such that
        \begin{align}
            -\zeta \leq \varrho \leq \zeta.
        \end{align}
        Then we have
        \begin{align}
            \left\|\varrho \right\|_2 \leq \left\|\zeta \right\|_2,
        \end{align}
        thus implying that $\varrho \in \mathscr{L}_{s\operatorname{HS}}(\mathscr{H})$.
    \end{lemma}
    }
    \Frame{
    \begin{remark}
        The matrix version of Lemma~\ref{lem:HS_operator_bound_norm} is known in the literature.
        For instance, it directly follows from Exercise~$3.2.7$ of~\cite{bhatia2009positive}.
    \end{remark}
    }
    \begin{proof}
        Set $\Psi \coloneqq \zeta - \varrho$ and $\Phi \coloneqq \zeta+\varrho$.
        The given hypothesis implies $\Psi, \Phi \in \mathscr{L}_{s}(\mathscr{H})_+$.
        We thus have
        \begin{align}
            \left\|\varrho \right\|_2^2
                &=\Tr\!\left[\varrho^2 \right] \\
                &= \frac{1}{4} \Tr\!\left[\left(\Phi-\Psi \right)^2 \right] \\
                &= \frac{1}{4} \Tr\!\left[\Phi^2+\Psi^2-\Phi \Psi-\Psi \Phi \right] \\
                &= \frac{1}{4} \left(\Tr\!\left[\Phi^2+\Psi^2 \right] - 2\Tr\!\left[\Phi \Psi \right]\right) \\
                &= \frac{1}{4} \left(\Tr\!\left[\Phi^2+\Psi^2 \right] - 2\Tr\!\left[\Phi^{\frac{1}{2}} \Psi \Phi^{\frac{1}{2}} \right]\right) \\
                &\leq \frac{1}{4} \left(\Tr\!\left[\Phi^2+\Psi^2 \right] + 2\Tr\!\left[\Phi^{\frac{1}{2}} \Psi \Phi^{\frac{1}{2}} \right]\right) \\
                &= \frac{1}{4} \Tr\!\left[\left(\Phi+\Psi \right)^2 \right] \\
                &=\Tr\!\left[\zeta^2 \right] \\
                &= \left\|\zeta \right\|_2^2,
        \end{align}
        concluding the proof.
    \end{proof}

    \Frame{
    \begin{theorem}\label{thm:int_hilbert_schmidt_valued_measure}
        If $\ell(E) \in \mathscr{L}_{s\operatorname{HS}}(\mathscr{H})_+$ for all $E \in \mathscr{B}(\mathscr{X})$, then for all $f \in L_\infty(\mathscr{X})$, we have
        \begin{align}
            \int_\mathscr{X} f(x) \ell(\dx) \in \mathscr{L}_{s\operatorname{HS}}(\mathscr{H}).
        \end{align}
        Furthermore, if $f$ is a non-negative function, then 
        \begin{align}
            \int_\mathscr{X} f(x) \ell(\dx) \in \mathscr{L}_{s\operatorname{HS}}(\mathscr{H})_+.
        \end{align}
        For $f, g  \in L_\infty(\mathscr{X})$, we have
        \begin{align}\label{eq:int_trace_HS_int}
            \Tr\!\left[\left(\int_\mathscr{X} f(x) \ell(\dx) \right)\left(\int_\mathscr{X} g(x') \ell(\dx') \right) \right] = \int_{\mathscr{X} \times \mathscr{X}} f(x)g(x') \Tr\!\left[\ell(\dx)\ell(\dx')\right].
        \end{align}
        Here $\Tr\!\left[m(\cdot) m(\cdot)\right]$ is the finite positive measure on $\mathscr{X} \times \mathscr{X}$ given by 
        \begin{align}
            \Tr\!\left[m(\cdot)m(\cdot)\right](E \times F)= \Tr\!\left[m(E)m(F)\right]
        \end{align}
        for all $E, F \in \mathscr{B}(\mathscr{X})$.
    \end{theorem}
    }
    
    \begin{proof}
        The arguments given in the proof of Theorem~\ref{thm:int_trace_class_valued_measure} work here verbatim with slight modifications.
        We provide the modified arguments in detail nonetheless, for the sake of readability.

        Suppose $\ell(E)\in \mathscr{L}_{s\operatorname{HS}}(\mathscr{H})_+$ for all $E \in \mathscr{B}(\mathscr{X})$, and let $f \in L_\infty(\mathscr{X})$.
        There exists a sequence $(f_n)_{n \in \mathds{N}}$ of simple measurable functions converging to $f$ uniformly so that
        \begin{align}\label{eq:int_wrt_povm_real_valued_maps_oper_norm_1}
            \int_\mathscr{X} f(x) \ell(\dx) = \lim_{n \to \infty} \int_{\mathscr{X}} f_n(x) \ell(\dx).
        \end{align}
        Again, we emphasize that the limit in~\eqref{eq:int_wrt_povm_real_valued_maps_oper_norm_1} exists with respect to the operator norm.
        It is easily seen that $\int_{\mathscr{X}} f_n(x) \ell(\dx) \in \mathscr{L}_{s\operatorname{HS}}(\mathscr{H})$ for all $n \in \mathds{N}$.
        We now show that the limit in~\eqref{eq:int_wrt_povm_real_valued_maps_oper_norm_1} exists in the Hilbert--Schmidt norm.

        It is easily seen that, for any simple measurable function $g$, we have 
        \begin{align}
            -\|g\|_\infty \ell(\mathscr{X}) \leq \int_{\mathscr{X}} g(x) \ell(\dx) \leq \|g\|_\infty \ell(\mathscr{X}).
        \end{align}
        By Lemma~\ref{lem:HS_operator_bound_norm}, we conclude that $\int_{\mathscr{X}} g(x) \ell(\dx)$ is a Hilbert--Schmidt operator, and it satisfies
        \begin{align}
            \left\| \int_{\mathscr{X}} g(x) \ell(\dx)  \right\|_2 \leq \|g\|_\infty \left\|\ell(\mathscr{X})\right\|_2.
        \end{align}
        The \emph{Hilbert--Schmidt norm semivariation} of $\ell$ is thus finite:
        \begin{align}
        \label{eq:hs_norm_variation_m}
            \left\| \ell \right\|_2 \coloneqq \sup \bigg\{\left\| \int_{\mathscr{X}} g(x) \ell(\dx) \right\|_2: \|g\|_\infty \leq 1, g\colon \mathscr{X}\to \mathds{R} \text{ simple} \bigg\} = \left\|\ell(\mathscr{X})\right\|_2 < \infty.
        \end{align}
        Consequently, the sequence $\left( \int_{\mathscr{X}} f_n(x) \ell(\dx) \right)_{n \in \mathds{N}}$ is Cauchy in the Hilbert--Schmidt norm, as seen by the following inequality:
        \begin{align}
            \left\| \int_{\mathscr{X}} f_n(x) \ell(\dx)-\int_{\mathscr{X}} f_{n'}(x) \ell(\dx) \right\|_2
                & \leq \left\| \ell \right\|_2 \left\| f_n - f_{n'} \right\|_\infty.
        \end{align}
        The norm inequality $\left\|\cdot \right\| \leq \left\| \cdot \right \|_2$ is known to hold on $\mathscr{L}_{s\operatorname{HS}}(\mathscr{H})$.
        This implies that the limit in~\eqref{eq:int_wrt_povm_real_valued_maps_oper_norm_1} with respect to the operator norm is the same as the limit of the sequence with respect to the Hilbert--Schmidt norm.
        Therefore, $\int_{\mathscr{X}} f(x) \ell(\dx) \in \mathscr{L}_{s\operatorname{HS}}(\mathscr{H})$.
        
        If $f$ is a non-negative function, then the sequence $\left( f_n \right)_{n \in \mathds{N}}$ can be chosen to be of non-negative simple measurable functions.
        In this case, we have $\mathscr{L}_{st}(\mathscr{H})_+$ and $\int_{\mathscr{X}} f_n(x) \ell(\dx) \in \mathscr{L}_{s\operatorname{HS}}(\mathscr{H})_+$ for all $n \in \mathds{N}$, which implies that $\int_{\mathscr{X}} f(x) \ell(\dx) \in \mathscr{L}_{s\operatorname{HS}}(\mathscr{H})_+$.

        Now, suppose $f, g \in L_{\infty}(\mathscr{X})$ are given.
        Let $\left( f_n \right)_{n \in \mathds{N}}$ and $\left( g_n \right)_{n \in \mathds{N}}$ be sequences of simple measurable functions uniformly converging to $f$ and $g$, respectively.
        We then have
        \begin{align}
            \Tr\!\left[  \left(\int_{\mathscr{X}} f_n(x) \ell(\dx) \right)\left(\int_{\mathscr{X}} g_n(x') \ell(\dx') \right)\right] =  \int_{\mathscr{X}} f_n(x)g_n(x') \Tr\!\left[\ell(\dx)\ell(\dx')\right].
        \end{align}
         Using the continuity of the inner product on $\mathscr{L}_{\operatorname{HS}}(\mathscr{H})$ given by the trace function and by taking the limit $n \to \infty$, we thus get  that
        \begin{align}
           \Tr\!\left[  \left(\int_{\mathscr{X}} f(x) \ell(\dx) \right)\left(\int_{\mathscr{X}} g(x') \ell(\dx') \right)\right] =  \int_{\mathscr{X}} f(x)g(x') \Tr\!\left[\ell(\dx)\ell(\dx')\right].
        \end{align}
        This concludes the proof.
    \end{proof}

\section{Bochner integral}   

\label{app:bochner_integral}

    We briefly review the theory of Bochner integrals and refer to~\cite[Chapter~1]{hytonen2016analysis} for a detailed overview of the topic.
    
    Let $\mathscr{X}$ be a measurable space equipped with a measure $\mu$.
    Let $(\mathds{B}, \left\|\cdot \right\|)$ be a Banach space over the complex field $\mathds{C}$.
    A 
    function $f\colon \mathscr{X} \to \mathds{B}$ is called simple if it is of the form $f \coloneqq \sum_{i=1}^n \mathbf{1}_{E_i} \otimes b_i$, where $b_i \in \mathds{B}$ and $E_i$ is a 
    {measurable} set of finite measure for all $1\leq i \leq n$.
    Here, we use the notation
    \begin{align}
        (g \otimes b)(x) \coloneqq g(x) b
    \end{align}
    for a function $g\colon \mathscr{X} \to \mathds{C}$ and $b \in \mathds{B}$.
    For a simple measurable function $f = \sum_{i=1}^n \mathbf{1}_{E_i} \otimes b_i$, define
    \begin{align}
        \int_{\mathscr{X}} f(x)  \, \mu(\dx)    \coloneqq \sum_{i=1}^n \mu(E_i) \, b_i.
    \end{align}
    A function $f\colon \mathscr{X} \to \mathds{B}$ is said to be strongly measurable if there exists a sequence $(f_n)_{n \in \mathds{N}}$ of simple measurable functions converging to $f$ $\mu$-almost everywhere.
    A strongly measurable function $f\colon \mathscr{X} \to \mathds{B}$ is said to be \emph{Bochner integrable} if there exists a sequence of simple measurable functions $f_n\colon \mathscr{X} \to \mathds{B}$  such that
    \begin{align}
        \lim_{n\to \infty}  \int_{\mathscr{X}} \left\|f(x)-f_n(x) \right\| \mu(\dx)  = 0.
    \end{align}
    In this case, the sequence $\Par{\int_{\mathscr{X}}  f_n(x)\mu(\dx)}_{n \in \mathds{N}}  $ is Cauchy in $\mathds{B}$.
    By completeness, the sequence is convergent in $\mathds{B}$, and the limit is called the \emph{Bochner integral} of $f$ with respect to $\mu$. 
    We write it using the notation
    \begin{align}
        \int_{\mathscr{X}} f(x)  \mu(\dx)  \coloneqq \lim_{n\to \infty} \int_{\mathscr{X}} f_n(x)  \mu(\dx) .
    \end{align}
    Here it is understood that the integration is with respect to the given measure $\mu$.
    If $f\colon \mathscr{X} \to \mathds{B}$ is Bochner integrable and $\Psi\colon \mathds{B} \to \mathds{B}^{\prime}$ is a bounded operator from the Banach space $\mathds{B}$ to another Banach space $\mathds{B}^{\prime}$ then the composition $\Psi \circ f\colon \mathscr{X} \to \mathds{B}^{\prime}$ is also Bochner integrable. 
    Moreover, we have
    \begin{align}\label{eq:bochner_int_bdd_map_action}
        \int_{\mathscr{X}} \Psi(f(x)) \mu(\dx) = \Psi\left(\int_{\mathscr{X}} f(x) \mu(\dx) \right).
    \end{align}
    See~\cite[Eq.~(1.2)]{hytonen2016analysis}.
    
    The special case relevant to our analysis occurs when $\mu(\mathscr{X})< \infty$ and $(\mathds{B}, \|\cdot \|)$ is a separable Banach space.
    This is because, in this case, every 
    strongly measurable function $f\colon \mathscr{X} \to \mathds{B}$ satisfying $\sup_{x \in \mathscr{X}} \|f(x)\| < \infty$ is Bochner integrable~\cite[Proposition~1.2.2]{hytonen2016analysis}. 
    In particular, the Bochner integral~\eqref{eq:average_state_integral} is well defined.

\section{Convolution property of the Gaussian score function}

    \label{app:gaussian_score}
    
    \Frame{
    \begin{proposition}
        \label{prop:gaussian_score_convolution}
        The function $S_\Sigma$ defined in~\eqref{eq:gaussian_score_function} satisfies the convolution property $S_\Sigma = P_\Sigma \star_\lambda P_\Sigma$, where $\lambda$ is the Lebesgue measure on $\mathds{R}^N$ and  $P_\Sigma\colon  \mathds{R}^{N} \times \mathds{R}^{N} \to \mathds{R}$ is the following symmetric and bounded map 
    \begin{align}
        P_\Sigma =  \left((2\pi)^{N} \det\left(\Sigma/4\right) \right)^{-1/4} S_{\Sigma/2}.
    \end{align}
    \end{proposition}
    }
    \begin{proof}
        For $x,y \in \mathds{R}^{N}$, we have
	\begin{align}
		&(S_{\Sigma/2} \star_\lambda S_{\Sigma/2}) (x,y) \nonumber \\ 
			&\coloneqq \int_{\mathds{R}^{N}} \ S_{\Sigma/2}(x, z) S_{\Sigma/2}(y, z) \dz \\
			&= \int_{\mathds{R}^{N}} \exp\!\left(- (z-x)^\intercal \Sigma^{-1} (z-x) \right) \exp\!\left(- (z-y)^\intercal \Sigma^{-1} (z-y) \right) \dz \\
			&= \int_{\mathds{R}^{N}} \exp\!\left(-(z-x)^\intercal \Sigma^{-1} (z-x) - (z-y)^\intercal \Sigma^{-1} (z-y) \right) \dz.
	\end{align}
        The exponent in the above integrand can be simplified to get
        \begin{align}
		&(S_{\Sigma/2} \star_\lambda S_{\Sigma/2}) (x,y) \nonumber \\ 
			&= \int_{\mathds{R}^{N}}  \exp\!\left(-\frac{1}{2} \left[   \left(z-\frac{1}{2}(x+y) \right)^\intercal \left(4\Sigma^{-1} \right) \left(z-\frac{1}{2}(x+y) \right)+ (x-y)^\intercal \Sigma^{-1}  (x-y) \right] \right) \dz \\
            &= S_{\Sigma}(x,y) \int_{\mathds{R}^{N}} \exp\!\left(-\frac{1}{2} \left[   \left(z-\frac{1}{2}(x+y) \right)^\intercal \left(4 \Sigma^{-1} \right) \left(z-\frac{1}{2}(x+y) \right) \right] \right) \dz \label{eq:convolution_1} \\
		&= S_{\Sigma}(x,y) \sqrt{(2\pi)^{N} \det \left(\Sigma/4\right)},
	\end{align}
	where we applied the Gaussian integral formula to get the last equality.
    \end{proof}

\section{The expected gain of a POVM is well defined}

\label{app:exp_gain_integral}

    Define a map $g: \mathscr{X} \to \mathscr{L}_{t}(\mathscr{H})$ by
    \begin{align}
         g(x) \coloneqq \rho_x \int_{\mathscr{X}} S(x,\hat{x}) m(\dhx).
     \end{align}
     It directly follows from Eq.~(7.6) of \cite{Mitter_etal_1984} that $g$ is Borel measurable.
     Therefore, the composite map $\Tr\left[g(\cdot)\right]\equiv G_{\mathscr{E}, S, m}(\cdot)$ is also measurable.
     This is because trace is a bounded linear map on $\mathscr{L}_t(\mathscr{H})$ and hence is Borel measurable. 
     Since $G_{\mathscr{E}, S, m}$ is a non-negative bounded map, it is integrable and whence the integral \eqref{eq:exp_gain} is well defined.

\end{document}